\documentclass[prb, twocolumn,superscriptaddress,showpacs]{revtex4}  
\usepackage{amssymb}
\usepackage[colorlinks,bookmarks=false,citecolor=blue,linkcolor=red,urlcolor=blue]{hyperref}
\usepackage{times}
\usepackage{amsmath}
\usepackage{bbm}
\usepackage{graphicx}
\usepackage{color}
\usepackage{enumerate}

\def\Tr{\text{Tr}}

\begin{document}

\author{Andreas M. L\"auchli}
\affiliation{Max Planck Institut f\"ur Physik komplexer Systeme, N\"othnitzerstrasse 38, D-01187 Dresden, Germany}

\author{Philipp Werner}
\affiliation{Theoretische Physik, ETH Zurich, 8093 Z{\"u}rich, Switzerland}

\title{Krylov-implementation of the hybridization expansion impurity solver\\ and application to 5-orbital models}

\date{\today}

\hyphenation{}

\begin{abstract}
We present an implementation of the hybridization expansion impurity solver which employs sparse matrix exact-diagonalization techniques to compute the time evolution of the local Hamiltonian. This method avoids computationally expensive matrix-matrix multiplications and becomes advantageous over the conventional implementation for models with 5 or more orbitals. In particular, this method will allow the systematic investigation of 7-orbital systems (lanthanide and actinide compounds) within single-site dynamical mean field theory. We illustrate the power and usefulness of our approach with dynamical mean field results for a 5-orbital model which captures some aspects of the physics of the iron based superconductors. 
\end{abstract}


\maketitle

\section{Introduction}

The development of efficient numerical methods to solve quantum impurity problems is an active research area. Demand for powerful and flexible impurity solvers is driven by the success of dynamical mean field theory (DMFT), which approximates Fermionic lattice problems by self-consistent solutions of appropriately defined quantum impurity models.\cite{Georges96} While impurity models are computationally more tractable than lattice models, the desire to include spatial correlations via cluster extensions\cite{Hettler98, Lichtenstein00, Kotliar01} or to treat  complicated interaction terms in realistic descriptions of multi-orbital systems results in considerable computational challenges. 

The multi-site or multi-orbital nature of the most relevant impurity models favors Monte Carlo methods. In this area, considerable progress has been achieved with the recent development of continuous-time or diagrammatic Monte Carlo techniques (CTQMC). The CTQMC algorithms come in two flavors. Weak coupling solvers\cite{Rombouts99, Rubtsov05, Gull08_ctaux} are based on an expansion of the partition function in powers of the interaction terms. This is the method of choice for large cluster calculations of relatively simple models (such as the one-band Hubbard model), because the computational effort scales as the cube of the system size. The complementary approach is based on an expansion of the partition function in the impurity-bath hybridization.\cite{Werner06} This so-called hybridization expansion technique treats the local interactions exactly and can be applied to a wide range of models, including the $t$-$J$ and Kondo-lattice model.\cite{Werner06Kondo, Haule07} However, since the Hilbert space of the local problem grows exponentially with the number of sites or orbitals, the computational effort scales exponentially, rather than cubically with system size. Nevertheless, the flexibility of the hybridization-expansion method and the information it can provide about the relevant states of the atomic system make it a desirable tool in particular for the DMFT study of transition metal oxides and actinide compounds. Here, we present an implementation of this method which enables the reliable simulation of models with up to seven orbitals on present-day compute clusters with $O(100)$ processors. 

The rest of this paper is organized as follows: Section \ref{sec:hybrid_expansion} provides a brief review of the hybridization expansion technique in the matrix formulation of Refs.~\onlinecite{Werner06Kondo, Haule07} and Section \ref{sec:krylov} discusses the new Krylov-based implementation. We demonstrate the accuracy and efficiency of the Krylov approach in Section \ref{sec:analysis}, and use it in Section \ref{sec:application} to compute phase diagrams for a ``toy model" of the pnictides (a five orbital model with almost degenerate bands and relatively large Hund coupling term). Section \ref{sec:conclusion} is a conclusion and outlook.

\section{Hybridization expansion in the general matrix formulation}
\label{sec:hybrid_expansion}
A quantum impurity model describes an atom or molecule embedded in some host material with which it can exchange electrons. The corresponding Hamiltonian $H=H_\text{loc}+H_\text{mix}+H_\text{bath}$ contains three terms: $H_\text{loc}=\sum_{\alpha,\beta}\epsilon^{\alpha,\beta}\psi_\alpha^\dagger\psi_\beta+\sum_{\alpha,\beta,\gamma,\delta}U^{\alpha, \beta, \gamma, \delta}\psi^\dagger_\alpha\psi^\dagger_\beta\psi_\gamma\psi_\delta$ describes the impurity (chemical potential, interaction and inter-site/orbital hopping terms), $H_\text{bath}=\sum_{\alpha,p} \epsilon_p^\alpha a^\dagger_{p,\alpha}a_{p,\alpha}$ a bath of non-interacting electrons whose parameters are fixed by the DMFT self-consistency,\cite{Georges96} and the hybridization term $H_\text{mix}=\sum_{\alpha, \alpha', p} (V^{\alpha, \alpha'}_{p} \psi^\dagger_\alpha a_{p,\alpha'} + h. c.)$ controls the exchange of electrons between the impurity and the bath. Diagrammatic Monte Carlo simulation relies on an expansion of the partition function $Z=\Tr[e^{-\beta H}]$ into a series of diagrams and the stochastic sampling of collections of these diagrams. For the hybridization expansion,\cite{Werner06, Werner06Kondo, Haule07} we split the Hamiltonian into two parts, $H_1=H_\text{loc}+H_\text{bath}$ and $H_2=H_\text{mix}$, and employ an interaction representation in which the time evolution of operators is given by $H_1$: $O(\tau)=e^{\tau H_1}O e^{-\tau H_1}$. In this interaction representation, the partition function can be expressed as a time ordered exponential, which is then expanded into powers of $H_2$,
\begin{eqnarray}
Z&=&\Tr \Big[e^{-\beta H_1} Te^{-\int_0^\beta d\tau H_2(\tau)} \Big]\nonumber\\
&=&\sum_{n=0}^\infty \int_0^\beta d\tau_1\ldots \int_{\tau_{n-1}}^\beta d\tau_n \Tr\Big[ e^{-(\beta-\tau_n)H_1}(-H_2) \ldots \nonumber \\
&&\hspace{25mm} \ldots e^{-(\tau_2-\tau_1)H_1}(-H_2)e^{-\tau_1H_1}\Big].
\label{Z_interaction_picture}
\end{eqnarray}
Equation~(\ref{Z_interaction_picture}) represents the partition function as a sum over all configurations $c=\{\tau_1<\ldots<\tau_n\}$, $n=0$, $1$, $\ldots$, $\tau_i\in[0,\beta)$ with weight
$w_c=\Tr[ e^{-(\beta-\tau_n)H_1}(-H_2)\ldots e^{-(\tau_2-\tau_1)H_1}(-H_2)e^{-\tau_1H_1}]d
\tau^n.$ 

After the expansion, the time evolution (given by $H_1$) no longer couples the impurity and the bath. It therefore becomes possible to integrate out the bath degrees of freedom analytically to obtain
\begin{widetext}
\begin{eqnarray}
w_{\tilde c}&=&Z_\text{bath}
\Tr_\text{loc} \Big[e^{-\beta H_\text{loc}} T 
\psi_{\alpha_{n}}(\tau_n)\psi^\dagger_{\alpha_n'}(\tau_{n}')
\ldots
\psi_{\alpha_1}(\tau_1)\psi^\dagger_{\alpha_1'}(\tau_1') 
\Big]\nonumber\\
&& \times \det M^{-1}(\{\tau_1, \alpha_1\},\ldots,\{\tau_{n},\alpha_n\}; \{\tau_1',\alpha_1'\},\ldots,\{\tau_{n}',\alpha_n'\}) (d\tau)^{2n}.
\label{weight_strong}
\end{eqnarray}
\end{widetext}
The configurations $\tilde c$ are now collections of $n$ time arguments $\tau_1<\ldots<\tau_n$ corresponding to annihilation operators with flavor indices $\alpha_1, \ldots,\alpha_n$ and $n$ time arguments $\tau_1'<\ldots<\tau_n'$ corresponding to creation operators with flavor indices $\alpha_1', \ldots, \alpha_n'$. The element $i,j$ of the matrix $M^{-1}$ is given by the hybridization function $F_{\alpha_i',\alpha_j}(\tau_i'-\tau_j)$, which is defined in terms of the hybridization parameters $V^{\alpha, \alpha'}_p$ and the bath energy levels $\epsilon^\alpha_p$.\cite{Werner06Kondo} Given the weights $w_{\tilde c}$, a stochastic sampling of all relevant configurations $\tilde c$ can be implemented using local updates such as the random insertion or removal of pairs of creation and annihilation operators.

For the present purpose, the important thing to note is that up to the irrelevant constant $Z_\text{bath}$ the weights consist of two factors: $\Tr_\text{loc}[\ldots]$ evaluates the imaginary-time evolution of the quantum impurity for a given sequence of hybridization events, while $\det M^{-1}$ gives the contribution of the bath degrees of freedom which have been integrated out. Using fast matrix updates, the determinant ratios for local updates can be computed in a time $O(n^2)$. The exponential scaling of the algorithm is due to the trace factor. With the exception of single-site multi-orbital systems with density-density interactions (for which the occupation number basis is an eigenbasis of $H_\text{loc}$ and thus the very efficient segment formulation\cite{Werner06} can be used), the exponential growth of $\dim(H_\text{loc})$ with number of sites or orbitals means that the simulation of large systems becomes computationally expensive.

The strategy proposed in Ref.~\onlinecite{Werner06Kondo} was to evaluate the trace in the eigenbasis of the local Hamiltonian. In this basis, the time evolution operators $e^{-\tau H_\text{loc}}$ become diagonal and can be evaluated easily. On the other hand, the operators $\psi$ and $\psi^\dagger$, which are sparse and simple in the occupation number basis, become complicated matrices in the eigenbasis of $H_\text{loc}$. To facilitate the task of multiplying these operator matrices it is important to order the eigenstates according to conserved quantum numbers as explained in Ref.~\onlinecite{Haule07}. 
The evaluation of the trace is then reduced to block matrix multiplications of the form
\begin{equation}
\sum_{\text{contr.} m} \Tr_m\Big[ \ldots (O)_{m'',m'}(e^{-(\tau'-\tau) H_\text{loc}})_{m'}(O)_{m',m}(e^{-\tau H_\text{loc}})_m\Big],
\end{equation}
where $O$ is either a creation or annihilation operator, $m$ denotes the index of the matrix block, and the sum runs over those sectors which are compatible with the operator sequence. With this technique, 3-orbital models or 4-site clusters can be simulated efficiently.\cite{Werner08nfl, Werner09, Haule07plaquette, Park08, Gull08plaquette} However, since the matrix blocks are dense and the largest blocks grow exponentially with system size, the simulation of 5-orbital models becomes already quite expensive and the simulation of 7-orbital models with 5, 6 or 7 electrons is only doable if the size of the blocks is severely truncated. 

In fact, one should distinguish two types of truncations: 
\begin{enumerate}[(i)]
\item{the truncation of the outer trace ($\sum_{\text{contr.} m}$) to those quantum number sectors or states which give the dominant contribution,
\label{ref:trunc_i}}
\item{the reduction of the size of the operator blocks $(O)_{m',m''}$ via elimination of high-energy states.
\label{ref:trunc_ii}}
\end{enumerate}
 The truncation of type (\ref{ref:trunc_i}) is harmless at low enough temperature, because it restricts the possible states at only a single point on the imaginary-time interval. On the other hand, truncations of the type (\ref{ref:trunc_ii}), if not done properly, can lead to systematic errors, whose effect will be hard to estimate in large systems, because the truncations are necessarily severe.

\section{Krylov-space method}
\label{sec:krylov}
As an alternative  strategy to evaluate the trace factor in Eq.~(\ref{weight_strong}) we propose to
\begin{enumerate}
\item adopt the occupation number basis, in which the $\psi$-operator matrices can easily be applied to any given state, and
in which the sparse nature of $H_\text{loc}$ can be exploited during the imaginary time evolutions by relying on efficient Krylov-space methods, 
\item to approximate the outer trace by a sum over the lowest energy states (i.e. truncation type~(\ref{ref:trunc_i}) introduced above). 
\end{enumerate}
This implementation involves only matrix-vector multiplications of the type $\psi^{(\dagger)} |v\rangle$ and 
$H_\text{loc}|v\rangle$, with sparse operators $\psi^{(\dagger)}$ and $H_\text{loc}$, and is thus doable in principle even for systems for which the multiplication of dense matrix blocks becomes prohibitively expensive, or for which the matrix blocks will not even fit into the memory anymore. Furthermore, no approximation of type (\ref{ref:trunc_ii}) is required, so that all excited states remain accessible at intermediate $\tau$ in the trace. The sparse nature of the hybridization operators is evident given the fact that they consist of creation and annihilation operators in the
occupation number basis.
$H_\text{loc}$ is sparse because the number of interaction terms is proportional to a small integer power of the number of orbitals, while the dimension of the matrix grows exponentially with the 
number of orbitals.

Our implementation is based on very efficient sparse matrix algorithms for the evaluation of matrix exponentials applied to 
a vector, i.e.~$\exp(-\tau H_\text{loc})|v\rangle$.~\cite{ParkLight87,HochbruckLubich97,MolerVanLoan03} These algorithms construct the Krylov space 
$\mathcal{K}_p(|v\rangle)=\text{span}\{|v\rangle, H_\text{loc}|v\rangle,  H_\text{loc}^2|v\rangle,\ldots, H_\text{loc}^p|v\rangle\}$ and
then approximate the full matrix exponential by the matrix exponential of the Hamiltonian projected onto the Krylov space $\mathcal{K}_p(|v\rangle)$.
In Ref.~\onlinecite{HochbruckLubich97} it has been shown rigorously that  these Krylov space algorithms converge rapidly as a function of $p$,
typically reaching convergence for very small iteration numbers $p\ll N_\text{dim}$, although the number of iterations depends on the 
time interval $\tau$.

Let us describe the algorithm for the trace evaluation in some more detail. First, during the initialization part of the simulation, the following
steps are required:
\begin{enumerate}
\item Obtain the low energy spectrum and eigenfunctions of $H_\text{loc}$ using (Band-)Lanczos or Davidson techniques, or alternatively diagonalize $H_\text{loc}$ completely
using full diagonalization techniques. The Band-Lanczos or Davidson techniques are needed to resolve the exact degeneracies of the eigenfunctions.
\item Decide which eigenstates of the spectrum are to be kept in the outer trace. It is important not to destroy the multiplet structure
of $H_\text{loc}$ when truncating the trace. The truncation criteria employed in our implementation are discussed in more detail in Sec.~\ref{sec:analysis}.
\end{enumerate}
Then, in the actual evaluation of a trace, we proceed as follows:
\begin{enumerate}
\setcounter{enumi}{2}
\item{Propagate a retained state in the trace up to the first hybridization event (forward and backward in time). Since the initial state is an eigenstate
of $H_\text{loc}$, this state is simply multiplied by an exponential factor for the first interval.}
\item{Apply the hybridization operator on the propagated state.}
\item{Propagate the current state up to the next hybridization event using the Krylov-space approach to the matrix exponential described above. The state
to be propagated is generically not an eigenstate of $H_\text{loc}$ anymore, so the Krylov space must be constructed up to a certain dimension.
The Krylov space size should not be kept fixed, but should be determined for each imaginary time interval according to a convergence criterion.
In the applications reported in the present paper the average Krylov space dimension is $\approx 2$.}
\item{Go back to step 4 if more hybridization operators are present.}
\item{Add the contribution of the propagated state to the trace.}
\item{Go back to step 3 until all retained states have been considered in the trace.}
\end{enumerate}

In the truncated trace approach it is important to measure the various local observables at $\tau=\beta/2$ in order to be least affected by the truncation 
of the trace at $\tau=0$ (and equivalently at $\tau=\beta$).

We conclude this section by illustrating the main advantage of the Krylov space method through a simple time complexity analysis of the algorithm.
Say we want to determine the trace of a given sequence of the hybridization 
operators $\psi$ and $\psi^\dagger$. According to the truncation (\ref{ref:trunc_i}) introduced above we perform the trace over $N_ \text{tr} \leq N_ \text{dim}$ states, where $N_ \text{dim}$ is the 
typical size of the impurity Hilbert space, which grows exponentially with the number of sites or orbitals contained in the ``impurity". Since there are 
$N_\text{hyb}$ hybridization events, the complexity of the application of the hybridization operators is 
$O(N_\text{hyb} \times   N_ \text{dim} \times N_ \text{tr})$. 
The imaginary time evolution on the other hand is nontrivial on $N_\text{interval}=N_\text{hyb}-1$ intervals. Based on Ref.~\onlinecite{HochbruckLubich97}, we assume a typical
number of iterations $N_\text{iters}\ll N_\text{dim}$ is needed to reach convergence for the imaginary time evolution of a single state $|v\rangle$ over
an interval length $\tau$. It follows that the complexity of the imaginary time evolution part is
$O(N_\text{tr} \times N_\text{interval} \times N_\text{iter} \times N_\text{dim})$ and the overall time complexity amounts to
\[ O(N_ \text{dim} \times N_ \text{tr} \times [ N_\text{hyb} + N_\text{interval} \times N_\text{iter} ]). \] 
In the worst case where we retain
all states in the trace $N_\text{tr}=N_\text{dim}$ the complexity scales as $N_\text{dim}^2$, but in the best case $N_\text{tr} = O(1)$ 
the time complexity is {\em linear} in $N_\text{dim}$.

In comparison the matrix formulation has a less favorable scaling with $N_\text{dim}$. In the case where we keep all states in the
trace the time complexity is $O(N_\text{interval}N_\text{dim}^3 )$ because of the expensive dense matrix-matrix multiplications,
whereas it is $O(N_\text{interval}\times N_\text{dim}^2 \times N_\text{tr} )$ for the truncated trace version.

While it is therefore obvious that in theory the Krylov space approach is the method of choice due to its superior $N_\text{dim}$ scaling,
in practice the precise numbers of $N_\text{tr}, N_\text{iter}$, and $N_\text{dim}$ will determine which one of the two
formulations performs better for a given problem with tractable Hilbert space size. In the following section we
address the performance and scaling of the two algorithmic formulations.

\section{Effect of truncation and efficiency}
\label{sec:analysis}

\subsection{Accuracy of the Krylov approach}

\begin{figure}[t]
\begin{center}
\includegraphics[angle=-90, width=0.9\columnwidth]{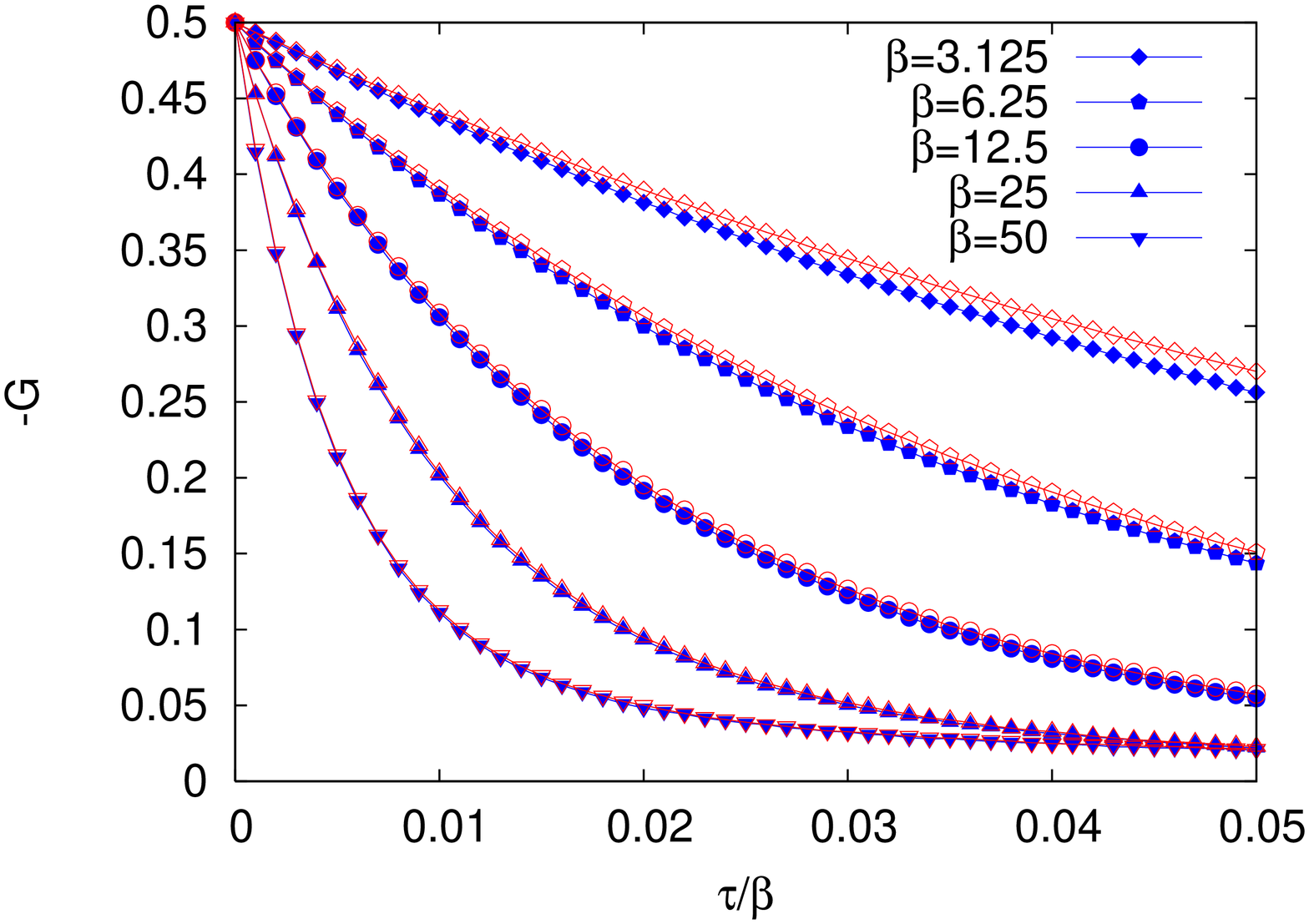}
\includegraphics[angle=-90, width=0.9\columnwidth]{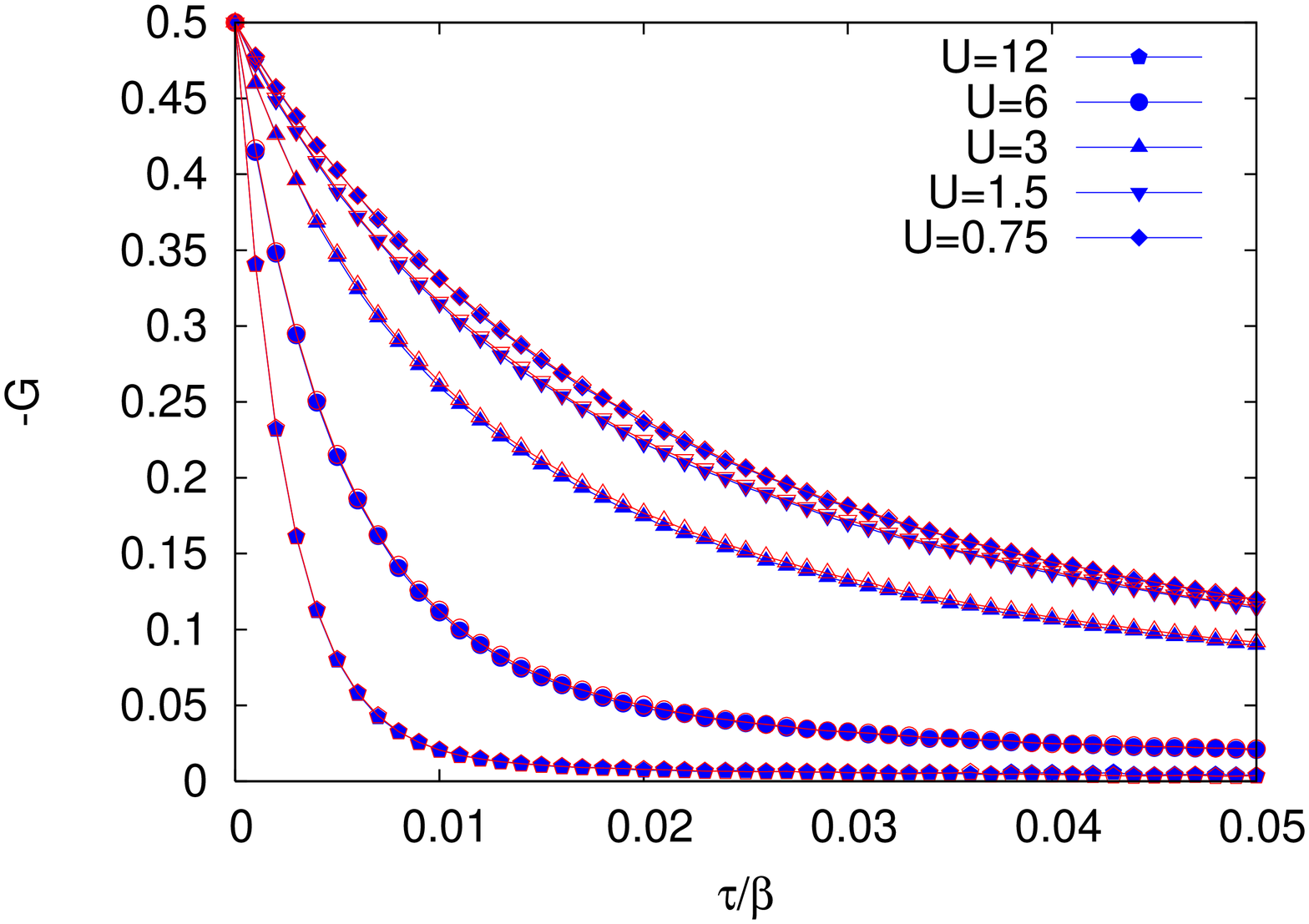}
\caption{(color online) Comparison between the Green functions of a 3-orbital model computed with the matrix method (open symbols, no truncation of the trace) and the Krylov method (full symbols, truncation of the trace to the lowest energy states) for different temperatures (top panel, $U=6$) and interaction strengths (bottom panel, $\beta=50$). The results become indistinguishable at temperatures which are $\lesssim 1\%$ of the bandwidth. 
}
\label{GFcomparison}
\end{center}
\end{figure}

Before trying to determine the system size for which the Krylov implementation outperforms 
 the matrix method, we  demonstrate the accuracy of the new approach. We consider multi-orbital models with a local Hamiltonian of the form
\begin{align}
H_\text{loc}&=-\sum_{a,\sigma} (\mu+\Delta_a)n_{a,\sigma} + \sum_{a} U n_{a,\uparrow} n_{a,\downarrow}\nonumber\\
&+\sum_{a>b,\sigma} \Big[U' n_{a,\sigma} n_{b,-\sigma} +  (U'-J) n_{a,\sigma}n_{b,\sigma}\Big]\nonumber\\
&-\sum_{a\ne b}J(\psi^\dagger_{a,\downarrow}\psi^\dagger_{b,\uparrow}\psi_{b,\downarrow}\psi_{a,\uparrow}
+ \psi^\dagger_{b,\uparrow}\psi^\dagger_{b,\downarrow}\psi_{a,\uparrow}\psi_{a,\downarrow} + h.c.)
\label{H_loc}
\end{align}
and rotationally invariant interactions ($U'=U-2J$, so the inter-orbital interactions are $U-2J$ for opposite spin and $U-3J$ for same spin).
Figure \ref{GFcomparison} compares the Green functions for a 3-orbital model with Hund coupling parameter $J=U/6$. The hybridization function is that of a non-interacting model with semi-circular density of states of bandwidth $4$ eV, and the chemical potential has been chosen such that the system is at half-filling ($\mu=\frac{5}{2}U-5J$). The crystal field splittings $\Delta_a$ are zero. We give the parameters $U$, $J$, $\mu$ and $\Delta$ in units of eV.

For three orbitals, both methods yield accurate results in a few CPU hours. The top panel of Fig.~(\ref{GFcomparison}) shows the measured Green functions for $U=6$, $J=U/6=1$ and different values of inverse temperature $\beta$, while the bottom panel compares the results for $\beta=50$ and different values of the interaction strength. The open symbols (red online) were computed with the matrix method without any truncations. These are thus exact results (Monte Carlo errors are much smaller than the symbol size) which may be used to test the accuracy of the Krylov approach. The Krylov results are plotted by full symbols (blue online). These data were computed with only the lowest energy states in the outer trace (four in this case, since the ground state of the half-filled model carries spin 3/2), which means that we use here the $O(1)$ approximation for $N_\text{tr}$. While deviations between the approximate and exact result are apparent at $\beta=3.125$, they become smaller as temperature is lowered and for $\beta \gtrsim 50$ can be considered negligible. The bottom panel shows that for $\beta=50$, essentially perfect agreement between the two methods is found for all relevant interaction strengths. 

These results can readily be understood from the scaling of the perturbation order with $U$ and $\beta$ in the hybridization expansion method.\cite{Werner06} The average perturbation order grows roughly linearly with the length $\beta$ of the imaginary time interval and decreases as interaction strength is increased. Since we restrict the system to the ground state at one point of the imaginary time interval ($\tau=0$), a larger number of hybridization events facilitates the relaxation into the true equilibrium distribution (measurements are performed at $\tau=\beta/2$). More importantly, the lower the temperature, the larger the overlap of this probability distribution with the ground state, i.e. the probability of the system being in the ground state at any given time becomes large. Thus, forcing the system into the ground state at  $\tau=0$ to compute the trace more efficiently has no severe effects at low enough temperature. Our data suggest that the truncation of $N_\text{tr}$ to the ground state vectors is legitimate for temperatures which are $\lesssim 1\%$ of the bandwidth ($4$~eV) and we will use this $O(1)$ truncation in all subsequent Krylov calculations.
We also note that the truncation of the trace does not seem to induce a sign problem for the multi-orbital
problems studied in the present work.

\begin{figure}[t]
\begin{center}
\includegraphics[angle=-90, width=0.9\columnwidth]{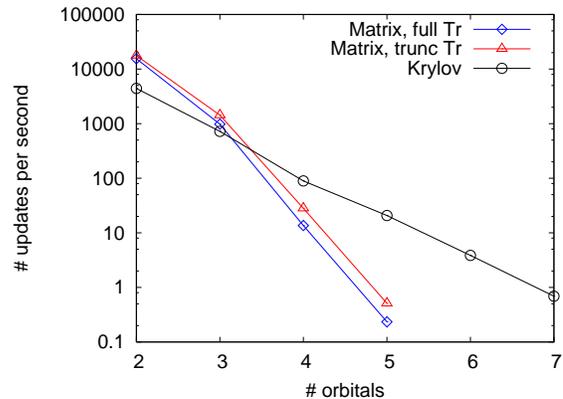}
\caption{(color online) Efficiency (number of updates per second) of the different implementations as a function of system size. The models are $n$-orbital impurity models with rotationally invariant Hund coupling at half-filling ($\mu=\mu_\text{half}$), 
$U=6$, $J/U=1/6$, $\beta=50$.
For the matrix method we show results without truncation (diamonds) and with truncation of the trace to the quantum number sector containing the ground state (triangles).
In the Krylov calculation, the trace is truncated to the lowest energy states. 
}
\label{scaling}
\end{center}
\end{figure}

\begin{figure}[t]
\begin{center}
\includegraphics[angle=-90, width=0.9\columnwidth]{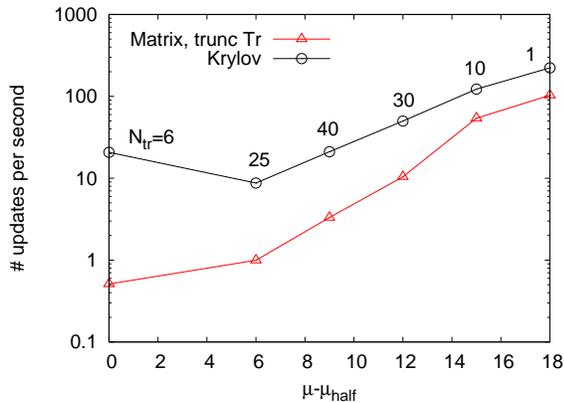}
\caption{(color online) Efficiency as a function of chemical potential for the 5 orbital model ($U=6$, $J/U=1/6$, $\beta=50$). The chemical potentials have been chosen such that the ground state of $H_\text{loc}$ has 5, 6, 7, 8, 9 and 10 electrons, respectively. The corresponding degeneracy $N_\text{tr}$ is plotted next to the Krylov data.
}
\label{scaling_mu}
\end{center}
\end{figure}

\subsection{Efficiency}

To compare the efficiency of the two implementations we plot in Fig.~\ref{scaling} the number of local updates per second for multi-orbital systems with $n=2,3,\ldots$ orbitals. A local update is either an insertion or a removal of a pair of $\psi$, $\psi^\dagger$ operators and involves the calculation of $\Tr_\text{loc}$. In our $n$-orbital models [Eq.~(\ref{H_loc})] each orbital interacts with every other through density-density, spin exchange and pair hopping terms. The intra-orbital repulsion is $U$, the Hund coupling parameter $J=U/6$, and the crystal field splittings are zero. We chose $U=6$, $\beta=50$ in all the calculations, and the hybridization function of the non-interacting model with semi-circular density of states of bandwidth $4$. The half-filling condition for these multi-orbital systems is $\mu_\text{half}=(n-\frac{1}{2})U-(n-1)\frac{5}{2}J$. The blue lines with diamonds show the results for the Matrix code without any truncation. For $n\ge3$, the evaluation of the trace becomes the bottleneck of the simulation and we observe an exponential decrease in the number of updates per second. The red lines with triangles show the result for the matrix code in which the trace is restricted to the sector $m$ containing the ground state (but without any truncation in the size of the blocks). 
The rather modest effect of the truncation is due to the fact that at half-filling the largest blocks cannot be discarded. 

The black lines with circles show the number of updates obtained with the Krylov-method. The curve still drops exponentially with increasing $n$, but the slope is smaller than in the matrix case, as expected from the scaling argument in the previous section. While the number of updates in the matrix implementation drops by about 4 orders of magnitude as $n$ is increased from 3 to 5, it drops only 2 orders of magnitude in the Krylov implementation. The more favorable scaling in the Krylov-case allows us to measure also $n=6$ and $n=7$ and as seen in Fig.~\ref{scaling}, the slope remains essentially unchanged. The time per update increases by about two orders of magnitude from $n=5$ to $n=7$. Given that the Krylov code allows the simulation of 5-orbital models (transition metal compounds) on a small number of processors, we therefore expect that this method will enable the controlled and accurate simulation of 7-orbital models (lanthanide and actinide compounds) on larger clusters with a few hundred processors.

\subsection{Effect of the ground state degeneracy}

The outer trace must at least contain all the ground state eigenvectors, and the ground state degeneracy of $H_\text{loc}$ depends on the model parameters. We therefore compare in Fig.~\ref{scaling_mu} the efficiency of the Matrix and Krylov implementations for the five orbital model at different values of the chemical potential (chosen such that the ground state lies in the $n_\text{tot}=5,6,\ldots, 10$ electron sector). The corresponding ground state degeneracies are 6, 25, 40, 30, 10 and 1. The increase in $N_\text{tr}$ from 6 to 25 leads to a slight decrease in the efficiency of the Krylov method if $\mu$ is increased from $\mu_\text{half}$ to $\mu_\text{half}+6$. For even larger $\mu$, the efficiency increases, because the relevant quantum sectors become smaller. This also explains why the advantage of the Krylov implementation over the Matrix implementation decreases as one moves away from half-filling. 

\begin{figure}[t]
\begin{center}
\includegraphics[angle=-90, width=0.9\columnwidth]{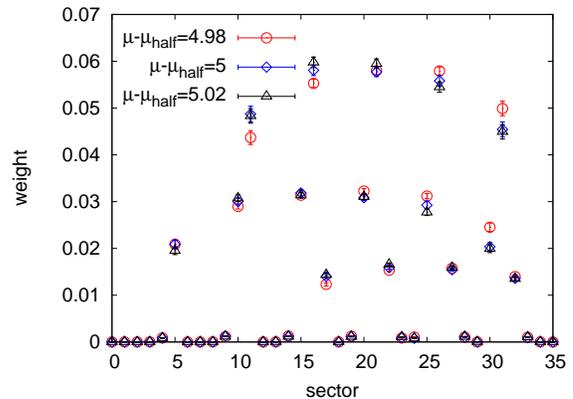}
\caption{(color online) Weight of the different $(n_\uparrow, n_\downarrow)$ quantum number sectors for $\mu-\mu_\text{half}=4.98$ ($N_\text{tr}=6$), $\mu-\mu_\text{half}=5$ ($N_\text{tr}=31$) and $\mu-\mu_\text{half}=5.02$ ($N_\text{tr}=25$).
}
\label{statistics}
\end{center}
\end{figure}

If the Krylov trace is restricted to ground state vectors, level crossings in $H_\text{loc}$ will typically lead to sudden changes in the number and types of states considered in $\text{Tr}_\text{loc}$, even in situations where the physical state of the system is not expected to change dramatically. However, as shown in Fig.~\ref{statistics}, this does not lead to inconsistencies if the temperature is sufficiently low. The figure plots the probability distribution of the different quantum number sectors, ordered as $(n_\uparrow=0, n_\downarrow=0), (n_\uparrow=1, n_\downarrow=0), \dots, (n_\uparrow=4, n_\downarrow=5), (n_\uparrow=5, n_\downarrow=5)$ from left to right, for the 5 orbital model with $U=6$, $J=U/6$ and $\mu=\mu_\text{half}+4.98$ (red circles, $N_\text{tr}=6$), $\mu=\mu_\text{half}+5$ (blue diamonds, $N_\text{tr}=31$) and $\mu=\mu_\text{half}+5.02$ (black triangles, $N_\text{tr}=25$) at $\beta=50$. All three trace calculations yield consistent distributions, and thus the same physical state. The (small) inaccuracies near level crossings could be further reduced by retaining all the states in a certain energy window above the ground state.

\section{Application}
\label{sec:application}
In this section we illustrate the usefulness and efficiency of the Krylov method with 
DMFT results for 5-orbital models with semi-circular density of states  of bandwidth 4.
We will consider the situation in which all bands are degenerate ($\Delta_a=0$, $a=1,\ldots, 5$) and a ``2+3" $e_g$-$t_{2g}$ crystal field splitting of magnitude 0.5, in which the doublet is shifted down ($\Delta_1=\Delta_2=0.5$, $\Delta_3=\Delta_4=\Delta_5=0$). All the calculations are for $\beta=50$ and require less than 50 CPU hours per iteration. 

\subsection{Orbitally degenerate case}

\begin{figure}[ht]
\begin{center}
\includegraphics[angle=-90, width=\columnwidth]{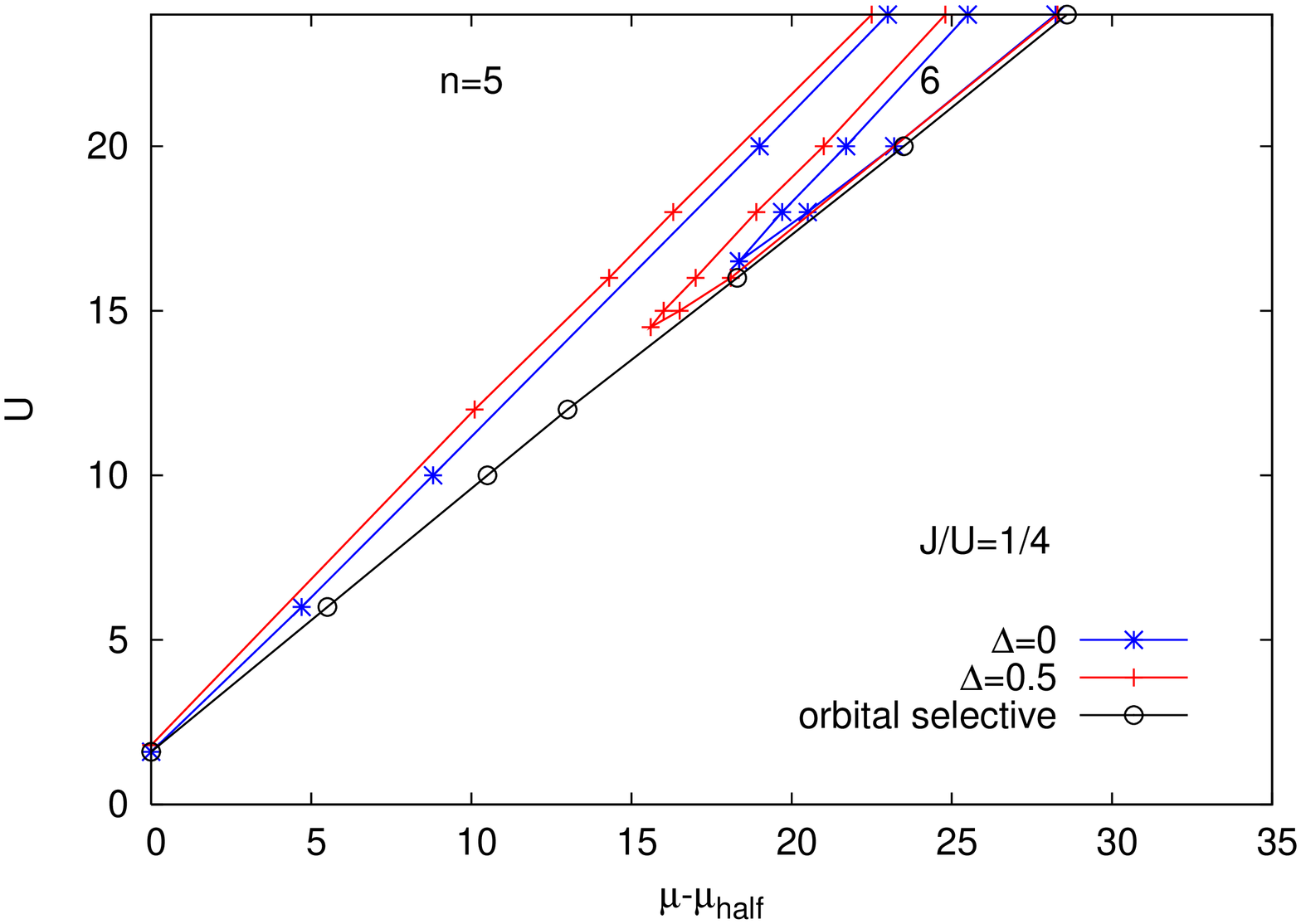}\\
\includegraphics[angle=-90, width=\columnwidth]{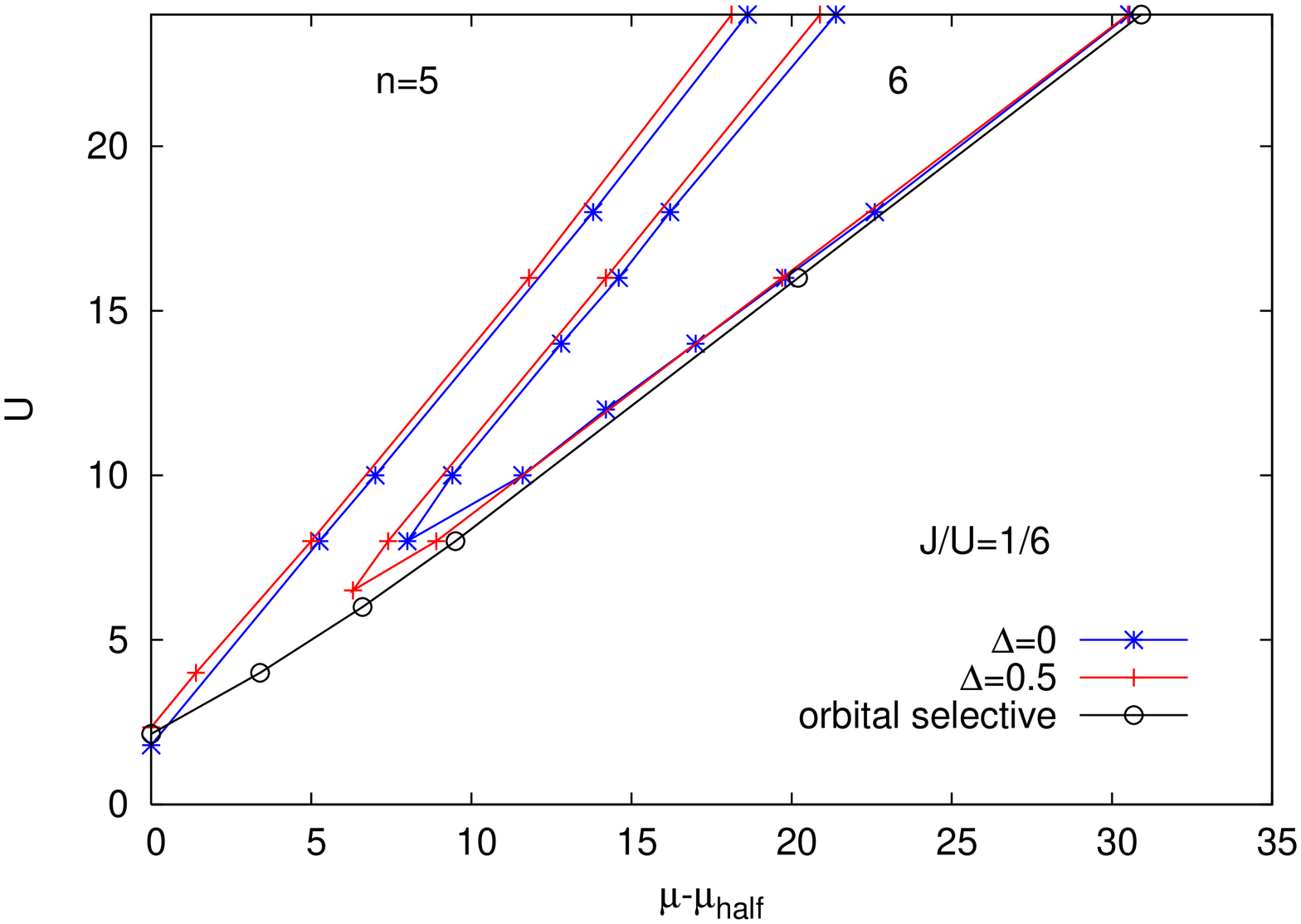}
\caption{(color online) Phasediagram of the 5-orbital model in the space of chemical potential and interaction strength for $J/U=1/4$ (top panel) and $J/U=1/6$ (bottom panel). The blue lines with stars show the $n=5$ and $n=6$ Mott lobes for the model without crystal field splitting. The red line with crosses shows the effect of a 2+3 crystal field splitting of magnitude $\Delta=0.5$ (2 orbitals shifted down) on the Mott lobes. Both Mott lobes are now contained in an orbital selective Mott phase (boundary marked with black circles) in which the three degenerate bands are insulating and half-filled, while the two degenerate bands are metallic. Error bars are on the order of the symbol size.
}
\label{phasediagram}
\end{center}
\end{figure}

Figure~\ref{phasediagram} shows the paramagnetic phase diagram in the space of chemical potential (relative to $\mu_\text{half}=4.5U-10J$) and interaction strength. The top panel is the result for $J/U=1/4$ and the bottom panel for $J/U=1/6$. We first discuss the orbitally symmetric case which corresponds to the blue lines with stars. The figure shows the Mott insulating lobes with $n=5$ and 6 electrons. Additional lobes with $n=7, \ldots, 9$ and a band insulating solution with $n=10$ also exist, but are not shown (computations near half-filling are the most challenging ones, because they involve quantum number sectors with high dimension). The Hund coupling $J$ is seen to stabilize the half-filled $n=5$ Mott lobe while pushing the critical interaction for the $n=6$ Mott lobe to larger values. The different widths of these lobes and their $J$-dependence can be understood from the $\mu$-dependence of the eigenstates of $H_\text{loc}$ as explained in the context of a 3-orbital calculation in Ref.~\onlinecite{Werner09}. A comparison of the 5-orbital result for $J/U=1/6$ to the lower panel of Fig. 2 in Ref.~\onlinecite{Werner09} and the 2-orbital calculations (Fig.~2) in Ref.~\onlinecite{Werner07crystal} furthermore shows the evolution of $U_c$ with increasing number of orbitals: in the 2-orbital model $U_c^\text{half filled}\approx 3.7$, in the 3-orbital model $U_c^\text{half filled}\approx 3$ and $U_c^\text{half filled+1}\approx 11$, and in the 5-orbital case we find $U_c^\text{half filled}\approx 2$ and $U_c^\text{half filled+1}\approx 8$.
Thus, in the presence of a Hund coupling $J=U/6$, the critical interaction strength for the Mott insulating phase decreases with increasing number of orbitals, in contrast to the situation for $J=0$.\cite{Ono03, Inaba05}

The very large interaction strengths required to study Mott physics away from half-filling are not a problem. An attractive feature of the hybridization expansion method is the fact that the relevant perturbation orders decrease with increasing interaction strength.\cite{Werner06} At $U=16$ and $\beta=50$, the $n=6$ Mott insulating solution for $J/U=1/6$ has average perturbation order $\approx 1.8$  per orbital and spin, i.e. $18$ in total.

\subsection{Crystal field splitting and orbital selective Mott transition}
 
We now consider the effect of shifting two orbitals down by $\Delta_1=\Delta_2=0.5$. The resulting phase diagram is plotted with crosses and red lines. The $n=5$ and 6 Mott lobes are little affected by the crystal field splitting. The value of $U_c$ is almost unchanged for $n=5$, and decreases by about 2 for $n=6$.  The width of the $n=6$ lobe is increased by about $\Delta$ and shrinks by a similar amount for $n=5$. Compared to the 2+1 splitting considered in Ref.~\onlinecite{Werner09} the stability of the $n=6$ lobe is not dramatically enhanced, because the insulating phase does not consist of half-filled and band-insulating solutions: the two degenerate bands ($a=1$, 2) accommodate 3 electrons. Larger effects on the stability of the $n=6$ lobe are expected for 4+1 or 3+1+1 splittings. 

A qualitative difference to the orbitally symmetric case is that the $n=5$ and 6 lobes are now embedded in an orbital selective Mott phase characterized by insulating, half-filled Green functions in the three degenerate bands ($a=3$, 4, 5) and metallic Green functions in the two lower lying bands ($a =1,2$). For $n=6$ electrons, the transition from the metallic into the orbital selective Mott phase takes place near $U\approx 12$ ($J/U=1/4$) and $U\approx 4.2$ ($J/U=1/6$), respectively. The metal-insulator transition in the three degenerate bands thus occurs at an interaction strength which is substantially smaller than the $U_c$ required to induce the Mott transition in the model without crystal field splitting. This finding is consistent with the enhanced stability of the half-filled Mott lobe in the 3-orbital model.\cite{Werner09}
 
\begin{figure}[t]
\begin{center}
\includegraphics[angle=-90, width=\columnwidth]{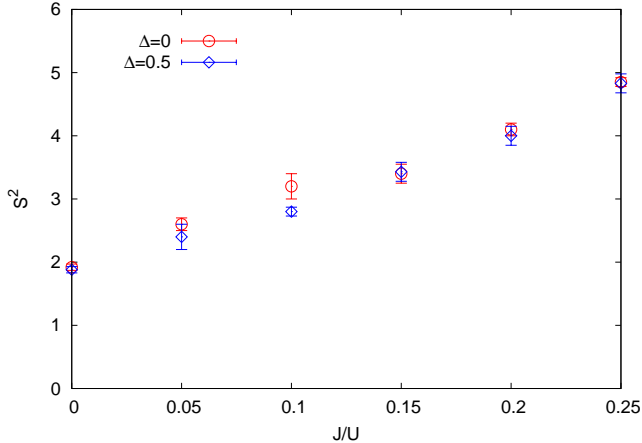}
\caption{(color online) Expectation value of $S^2$ as a function of $J/U$ for the five orbital model with $n=6$, $U=2$ and $\beta=50$. Circles show the result for degenerate orbitals, diamonds for a crystal field splitting $\Delta=0.5$ which shifts two orbitals down.}
\label{Stot}
\end{center}
\end{figure}  

\subsection{Total spin}

The Krylov implementation retains the attractive features of the hybridization expansion, such as the ability to measure the relevance of different states in the Hilbert space of $H_\text{loc}$ (Fig.~\ref{statistics}). As a practical application we plot in Fig.~\ref{Stot} the expectation value of the total spin squared, $S^2$, as a function of Hund coupling $J/U$ in the weakly correlated metallic phase ($U=2$, $\beta=50$) with 6 electrons. Results for degenerate bands and for a 2+3 crystal field splitting $\Delta=0.5$ are shown. 
The atomic ground states for $\Delta=0$ (0.5) correspond to $S^2=0$, $2$, $6$ ($0$, $2$) for $J/U=0$, $S^2=6$ ($2$) for $J/U=0.05$ and $S^2=6$ ($6$) for $J/U\ge 0.1$. The atomic picture is however not a good reference in the parameter regime considered here.  
We conclude from Fig.~\ref{Stot} that in the moderately correlated metallic phase the lower spin states have appreciable weight, and the effect of the crystal field splitting is small. No dramatic increase in $S^2$ is observed as $J/U$ is increased from 0 to 0.05, and the crystal field splitting of $\Delta=0.5$ leads to no significant reduction in $S^2$, even at $J=0$. 

\subsection{Implications for pnictides}

The toy model considered here captures some aspects of the iron-based high temperature superconductors. A minimal description of these materials seems to require all five $d$-bands,\cite{Kuroki08} and the bandwidth of 4 eV adopted here is consistent with the band structure obtained from density functional theory. Crystal field splittings appear to be small ($\Delta\approx 0.2$-$0.5$), although (due to the tetrahedral coordination) not of the simple 2+3 type considered in Fig.~\ref{phasediagram}.\cite{Kuroki08, HauleNJP} No consensus has yet emerged about the interaction parameters $U$ and $J$ and the role of correlations. Some authors argue that $U$ should be quite large and the material close to a Mott transition \cite{HauleNJP} or to an orbitally selective Mott state. \cite{Craco08,DeMedici09} Other theoretical studies, however, adopt small interaction parameters, $U\approx 2$, and $J=0.2$-$0.6$.\cite{Kuroki08,Kuroki09} 

Our phase diagrams for $J/U=1/6$ and $J/U=1/4$ in Fig.~\ref{phasediagram} can provide some information about the role of correlations and crystal field splittings, and the relevance of Mott physics. In particular, we see that the solution with $n=6$ and $U=2$ is far away from the Mott lobe in the orbitally symmetric case, especially if the Hund coupling is large. On the other hand, our results for the 2+3 splitting indicate that even relatively small crystal field splittings can have a substantial effect on the  phase diagram and lead to the opening of a gap in some bands at $U$ much below the $U_c$ for the fully gapped phase. Correlations in pnictides may thus be relevant in the sense that the materials are not too far from an orbital selective Mott state. But for such a scenario, the Hund coupling parameter $J$ would have to be rather small: for $J/U=1/4$, $U=2$ is a factor of 6 below the critical value for the orbital selective transition and for $J/U=1/6$ it is still a factor of 2 below. A moderately correlated metallic state seems more consistent with the phase diagram of our simple toy model. Figure~\ref{Stot}, on the other hand, indicates that the total spin in the realistic parameter regime is not particularly sensitive to the values of $J$ and $\Delta$. At $U=2$, the spin 2 states start to dominate only for $J>0.6$. However, since details of the band structure appear to affect the properties of iron based superconductors in profound ways,\cite{Kuroki09} more realistic calculations within the LDA+DMFT framework would be required to settle these issues. \\

\section{Conclusions} 
\label{sec:conclusion}
We have presented an implementation of the hybridization expansion impurity solver which makes use of {\em sparse-matrix} exact diagonalization techniques to evaluate the weight of Monte Carlo configurations. This method, while still scaling exponentially with system size, enables a much more efficient simulation of large multi-orbital problems than the established matrix formulation. In the new approach, the trace is restricted to a small number of states (typically the lowest energy eigenstates of the Hamiltonian) but no other truncations or approximations are necessary during the time evolution of these states.\footnote{This idea could also be adapted to the matrix method, where an implementation based on matrix-vector multiplications rather than matrix-matrix multiplications is expected to be favorable once the dimension of the relevant matrix blocks becomes large.} We have demonstrated that the restriction of the outer trace leads to negligible systematic errors at low temperature. Therefore, the Krylov method provides a controlled and efficient implementation of the hybridization expansion approach which enables the DMFT study of transition metal and actinide compounds with realistic interactions. 

\acknowledgments

We thank H. Aoki, R. Valenti, I. Eremin and E. Gull for helpful discussions.  PW was supported by SNF grant PP002-118866.
The simulations have been performed at the ZIH (TU Dresden), CSCS Manno and on the Brutus cluster at ETH Z\"urich, using some of 
the ALPS  libraries.\cite{alps}

\end{document}